\pdfoutput=1
\documentclass[journal=jacsat,manuscript=article]{achemso}   %%% Original
\usepackage{amsfonts}
\usepackage{amssymb}
\usepackage{graphicx}
\usepackage{amsmath}
\usepackage{tikz}
\usepackage{adjustbox}
\usepackage[normalem]{ulem}
\usetikzlibrary{shapes.geometric, arrows}
\usepackage[english]{babel}
\usepackage{color}
\usepackage[version=3]{mhchem} % Formula subscripts using \ce{}
\usepackage{natbib}
\usepackage{url}
\usepackage[colorlinks,citecolor=black,urlcolor=blue,linkcolor=black]{hyperref} % optional

\newcommand{\olr}[1]{{\color{red}{}}}% \sout{#1}}}

\newcommand{\osvp}[1]{{\color{blue}{}}}% \sout{#1}}}

\author{Sebastian V. Pios}
\affiliation{Zhejiang Laboratory, Hangzhou 311100, China}
\author{Jiaji Zhang} %\,\orcidlink{0000-0003-2978-274X}}
\affiliation{Zhejiang Laboratory, Hangzhou 311100, China}
\author{Maxim F. Gelin}
\affiliation{School of Science, Hangzhou Dianzi University, Hangzhou 310018, China}
\author{Hong-Guang Duan}
\affiliation{Department of Physics, School of Physical Science and Technology, Ningbo University, Ningbo, 315211, P.R. China.}
\author{Lipeng Chen}
\affiliation{Zhejiang Laboratory, Hangzhou 311100, China}
\email{chenlp@zhejianglab.com}

\title{Tracking the Electron Density Changes in Excited States - A Computational Study on Pyrazine}

\begin{document}

%\section{TOC Graphic}
%\begin{figure}
%    \centering
%   \includegraphics{toc.eps}   
%\end{figure}

%\begin{tocentry}
%  \includegraphics{toc.eps}
%\end{tocentry}

\begin{abstract}
The development of X-ray free-electron lasers (XFELs) has enabled ultrafast X-ray diffraction (XRD) experiments, which are capable of resolving electronic/vibrational transitions and structural changes in molecules, or capturing molecular movies.
While time-resolved XRD has received increasing attention, the extraction of information content from signals is challenging and requires theoretical support. In this work, we combined X-ray scattering theory and trajectory surface hopping approach to resolve dynamical changes in the electronic structure of photo-excited molecules by studying time evolution of electron density changes between electronic excited states and ground state. Using pyrazine molecule as an example, we show that key features of reaction pathways can be identified, enabling the capture of structural changes associated with electronic transitions for a photo-excited molecule.
\end{abstract}

%\section{TOC Graphic}
%\begin{figure}
%    \centering
%   \includegraphics{toc.eps}   
%\end{figure}

Modern techniques of ultrafast laser spectroscopy and time-resolved X-ray (electron) diffraction, which are capable of achieving femtosecond  temporal resolution and subangstrom spatial resolution, provide excellent probes of chemical reactions in various molecular systems. \cite{Mukamel2017Review, UED1, Young_2018, mol_movies_Wang2024} 
Thanks to the development of X-ray free-electron lasers (XFELs),\cite{Xray1}  ultrafast X-ray scattering experiments together with ultrafast spectroscopy techniques enable us to resolve directly electronic/vibrational transitions and structural changes of molecules, capturing the molecular movies of chemical reactions. With its high spatiotemporal resolution and signal-to-noise ratio, ultrafast X-ray diffraction has been used to study a variety of photophysical and photochemical processes, such as laser-induced molecular alignment,\cite{Xray_isolated_kuepper} electrocyclic ring-opening dynamics,\cite{Imaging_minitti} molecular vibrational excitation,\cite{stankus2019ultrafast} and excited-state charge transfer. \cite{ultrafast_xray_yong2021}

However, extracting the information of molecular structural changes from ultrafast X-ray diffraction data is not straightforward, theoretical support is usually required. A widely used model for the simulation of X-ray diffraction patterns is the so-called independent atom model (IAM),\cite{als2011elements} where the electron density is approximated by a sum of the electron density of each atom centered at its nuclear coordinate \textbf{R}. While the IAM offers a reasonable way to obtain approximated diffraction patterns in many circumstances, it does not distinguish between
different electronic states and does not consider the distortion of the electron density in chemical reactions. \cite{prince2004international} The latter issue can be partially alleviated by introducing generalized form factors specific for distorted charge distributions. \cite{generalized_xrd_factors,stewart_diatomic_1975} 

In order to circumvent aforementioned difficulties of the IAM, \textit{ab initio} simulation protocols have been developed to obtain an accurate description of time-resolved X-ray diffraction patterns.
The \textit{ab initio} X-ray diffraction (AIXRD) simulation routine pioneered by Kirrander and coworkers has been employed to investigate molecular geometrical changes in the electronic ground- and excited-states.\cite{Kirrander2014, Kirrander2016JCTC} Combining quantum dynamics methods and \textit{ab initio} quantum chemistry, time-resolved X-ray diffraction pattern for the 
isomerization of azobenzene was simulated, yielding movie of the conical intersection dynamics. \cite{MukamelPNAS2021} \textit{Ab initio} simulations of the inelastic scattering signal of linear cyanotetracetylene molecule provide image of electron-density fluctuations, revealing various scattering pathways that dominate the signal. \cite{mukamel_xrd2}

In this study, we introduce an approach that combines X-ray scattering theory and trajectory surface hopping method to simulate time-resolved X-ray diffraction patterns of polyatomic molecules in the gas phase. The proposed approach is illustrated by tracking the electron density changes in electronic excited states using pyrazine molecule as an example. The photophysics of pyrazine has been extensively studied by \textit{ab initio} quantum-chemistry calculations,\cite{woywod_pyrazine,gatti_pyrazine} dynamics simulations,\cite{xie_pyrazine,CLP} femtosecond photoelectron spectroscopy measurements,\cite{radlof_pyr_phelsp,suzuki2010time,suzuki2016full} and calculations of nonlinear spectroscopic signals.\cite{OTFDW1,Xiang2D,Skw20a,LP21,OTFDW2,OTFDW4} Furthermore, transient X-ray absorption spectra of pyrazine were evaluated by employing the multiconfiguration time-dependent Hartree (MCTDH) method.\cite{Freibert_2021,freibert_2024}

%%% Basic Theory
%%% XRD
The X-ray scattering cross-section of a system with \textit{N}\textsubscript{el} electrons can be expressed using Fermi's golden rule as \cite{james_optical_principles1982, schuelke_electron_dyn2007}
\begin{equation}\label{eq:elastic_scattering_manuscript}
    \frac{\mathrm{d}I}{\mathrm{d}\Omega} \biggr|_{\alpha_0} \propto \sum_{f} \left( \frac{\omega_{f}}{\omega_{0}}\right) |\langle \Psi_{\alpha_f} | e^{i\textbf{q}\textbf{r}} | \Psi_{\alpha_{0}}\rangle |^{2} .
\end{equation}
Here, $\textbf{r}$ is the electron coordinate, $\omega_{f}$ and $\omega_{0}$ are the frequencies of scattered and emitted lights, respectively. $\Psi_{\alpha_f}$ and $\Psi_{\alpha_0}$ are wave functions for the final and initial electronic states, respectively. For the coherent scattering, where $\alpha_f=\alpha_0=\alpha$, one can define the molecular form factor $f^{\alpha}(\textbf{q})$ as
\begin{equation}
f^{\alpha}(\textbf{q})=\langle\Psi_{\alpha}|e^{i\textbf{q}\textbf{r}}|\Psi_{\alpha}\rangle,
\end{equation}
where $\textbf{q}$ is the momentum transfer vector, which is defined as the difference between the incident and scattered wave vectors. In this work, we focus on the elastic scattering where the molecular form factor is assumed to be instantaneous, depending only on the current geometry and the distribution of the electronic state.\cite{Kirrander2014} The molecular form factor is obtained by the Fourier transformation
\begin{equation}\label{eq:form_factor_tot_manuscript}
    f^{\alpha}(\textbf{q}, t) = \int d\textbf{r} \, \rho^{\alpha} (\textbf{r},t) e^{i\textbf{qr}},
\end{equation}
where $\rho^{\alpha}(\textbf{r},t)$ is the electron density of the state $\alpha$ at time $t$. To visualize the electron density changes in the electronic excited state $S_n$ relative to the electronic ground state $S_0$, we introduce a quantity called ``molecular form factor difference", which is defined as 
\begin{equation}\label{eq:diff_form_factor_manuscript}
f^{S_{n}}_{diff} (\textbf{q}, t) = \int d\textbf{r} ~ \left[ \rho^{S_{n}} (\textbf{r}, t) - \rho^{S_0} (\textbf{r}, t) \right] e^{i\textbf{qr}}.
\end{equation}
This allows for a closer inspection of characteristic features of electron density changes without losing the phase information. The averaged molecular form factor difference is obtained as a weighted sum over all the excited states 
\begin{equation}
f_{diff}(\textbf{q},t)=\sum_{S_n}p_{S_n}(t)f_{diff}^{S_n}(\textbf{q},t),
\end{equation}
with $p_{S_n}(t)$ being the population of the electronic excited state $S_n$.

%%% NAD-QC MD
We use the trajectory surface hopping (TSH) approach \cite{OTF2,OTF3} to simulate the nonadiabatic dynamics of pyrazine where the electron density is obtained on-the-fly from the electronic structure program package. In TSH, trajectories are propagated in time on a single potential energy surface (PES) and nonadiabatic events are treated via instantaneous switches, so-called ``hops". Many protocols have been developed to treat nonadiabatic transitions, each with its respective strengths and weaknesses.\cite{OTF1,OTF2,mai2020molecular} In this work, we employ a protocol based on the Landau-Zener formalism (see the Supporting Information for details),\cite{LZM1,LZM2} which does not suffer from overcoherence problem. In the Landau-Zener formalism, the transition probability is estimated by the energy difference between adjacent adiabatic electronic states and its second derivative with respect to time, thus avoiding the calculation of possible divergent nonadiabatic couplings. By only allowing transitions between adjacent adiabatic states, the number of excited states that need to be calculated can be decreased significantly.\cite{pragmatic_lz}

%%% RESULTS AND DISCUSSION
%\section{Results and discussion}
%
\begin{figure}
    \centering
    \includegraphics[width=0.7\textwidth]{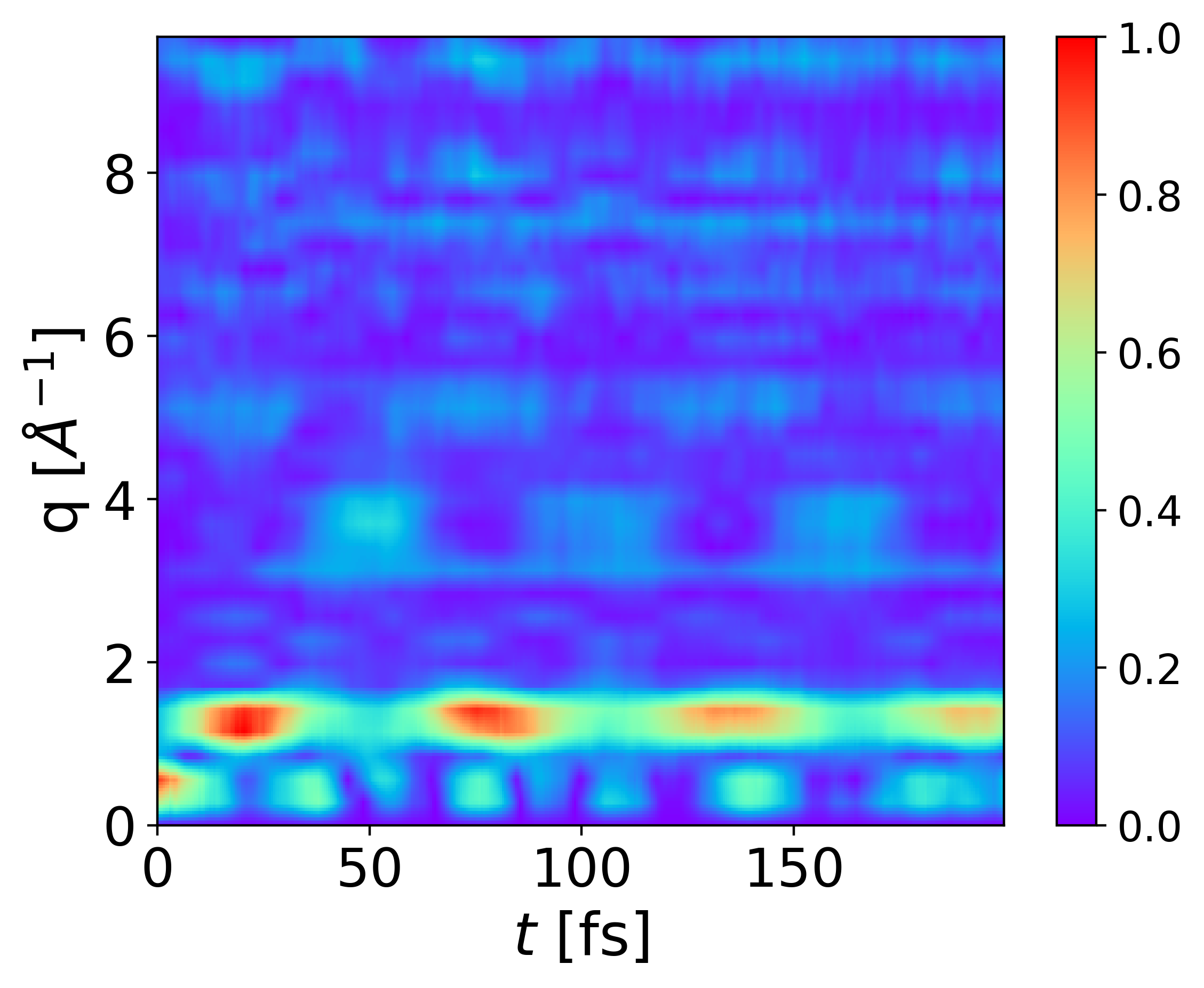}
    \caption{Absolute value of the averaged molecular form factor difference, $|f_{diff}(\mathrm{\textbf{q}}, t)|$, as a function of the momentum transfer q and $t$. 
    }
    \label{fig:abs_time}
\end{figure}

Figure~\ref{fig:abs_time} shows the absolute value of the averaged molecular form factor difference, $|f_{diff}(\mathrm{\textbf{q}},t)|$, as a function of the momentum transfer q and $t$ for pyrazine molecule (the real and imaginary components of $f_{diff}(\textbf{q},t)$ are given in Figures S1 and S2 of the Supporting Information, respectively). In the simulation, a X-ray wavelength of  1.3~\AA\space is used. Five distinct features at q $\approx$ 0.6, 1.3, 3.6, 5.0 and 8.8~\AA\textsuperscript{-1} can be observed, which exhibit oscillations that reflect key processes characteristic for photo-excited pyrazine. At \textit{t} = 0~fs, the signal is dominated by the features at q $\approx$ 0.6~\AA\textsuperscript{-1} with minor contributions at other q. From \textit{t} = 0~fs to \textit{t} = 25~fs, one can clearly observe the shift of the signal towards q $\approx$ 1.3~\AA\textsuperscript{-1}, which reflects the fast population transfer from the initially populated B\textsubscript{2u} state to the lower-lying A\textsubscript{u} and B\textsubscript{3u} states. The signal at q $\approx$ 0.6~\AA\textsuperscript{-1} exhibits the recurrence pattern with a period of $\approx$ 33~fs, which can be attributed to the tuning mode \textit{Q}\textsubscript{1}. This recurrence pattern can be also seen at q $\approx$ 8.8~\AA\textsuperscript{-1}, but with much weaker intensity. The signal at both q $\approx$ 0.6~\AA\textsuperscript{-1} and q $\approx$ 1.3~\AA\textsuperscript{-1} also shows recurrence patterns at \textit{t} $\approx$ 80, 140 and 195~fs, which correlates with the population revival of the B\textsubscript{2u} state. These recurrences are primarily caused by the modulation of the elctron density in the B\textsubscript{3u}/A\textsubscript{u} states by the vibrational $\nu_{\mathrm{6a}}$ mode, which, owning to substantial anharmonicity, has a period around 50-60~fs. 
The beating caused by this mode is also manifested in the evolution of off-diagonal peaks in electronic 2D spectra in pyrazine.\cite{kewei_beating_maps_2023}
The features at q $\approx$ 3.6~\AA\textsuperscript{-1} and q $\approx$ 5.0~\AA\textsuperscript{-1} are characteristic for  the  A\textsubscript{u} state and  B\textsubscript{3u} state, respectively.   

\begin{figure}
    \centering
    \includegraphics[width=\textwidth]{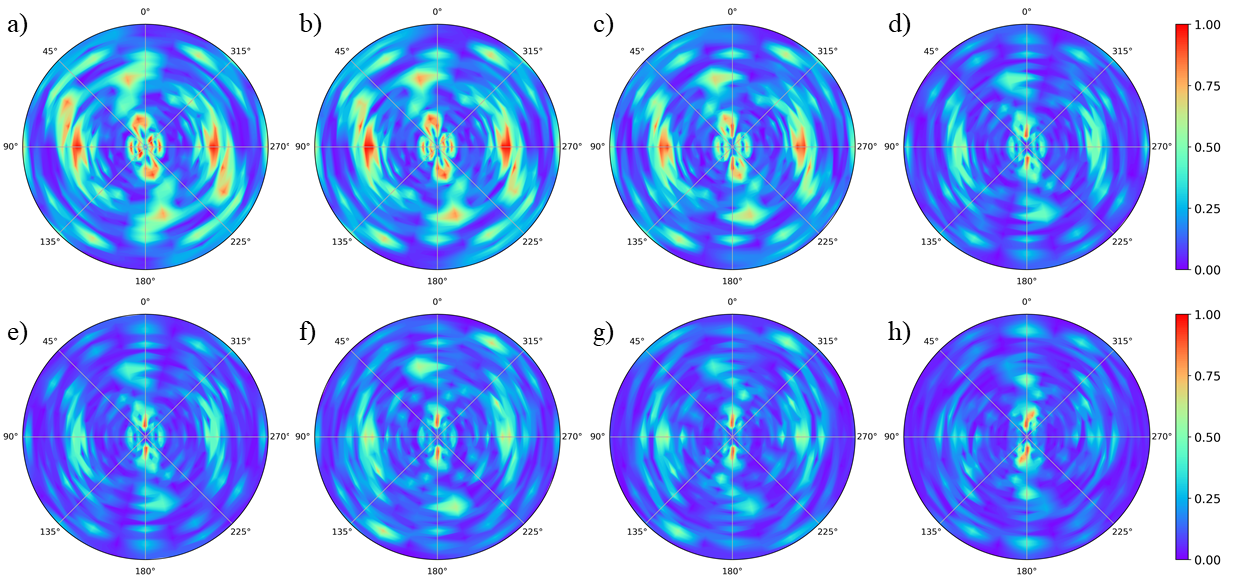}
    \caption{Absolute value of the averaged molecular form factor difference, $|f_{diff}(\mathrm{\textbf{q}}, t)|$, as a function of the momentum transfer q and the azimuthal angle $\phi$ at times \textit{t} = 0.0, 5.0, 10.0, 25.0, 32.0, 65.0, 80.0, 120.0~fs (a-h respectively). The signals are rescaled for improved readability.}
    \label{fig:avg_f_xy_dens_form_abs}
\end{figure}

To gain more structural details from the electron density changes, we present the absolute value of the averaged molecular form factor difference, $|f_{diff}(\textbf{q},t)|$, as a function of the momentum transfer q and azimuthal angle $\phi$ at times \textit{t} = 0.0, 5.0, 10.0, 25.0, 32.0, 65.0, 80.0, 120.0~fs in Fig.~\ref{fig:avg_f_xy_dens_form_abs}. The real and imaginary components of $f_{diff}(\textbf{q},t)$ as a function of the momentum transfer q and azimuthal angle $\phi$ are displayed in Figures~S3 and S4 of the Supporting Information, respectively. The molecule is placed in the xy-plane with N atoms aligned vertically, and the X-ray pulse is polarized along the z-axis. The signals show near-centrosymmetric patterns with a 2-fold symmetry axis, revealing that the photoexcited pyrazine does maintain its planar structure throughout the simulation time. From \textit{t} = 0~fs to \textit{t} = 10~fs, one observes significant changes in the electron density along the C-C bonds, as illustrated by the intensity changes of signals at azimuthal angles $\phi \approx 90^{\circ}$ and $\phi \approx 270^{\circ}$.
While the signal along the C-C bonds decreases in intensity with time, the intensity of the signal around the N atoms and C-N bonds increases, which reflects the fact that the wavepacket transfers to low-lying electronic states. Signals around the C-H bonds are weak throughout the simulation time, indicating that the electron density changes have minor contributions from those bonds.          

To determine the relative positions of the signal, we analyze the initial and final orbitals for each electronic excited state (see Table~S1 and Figure~S5 of the Supporting Information). The bright B\textsubscript{2u} state consists of multiple transitions, the main of which comes from a transition between the occupied 1b\textsubscript{1g} and the virtual 2b\textsubscript{3u} molecular orbitals. While those orbitals overlap along the C-C bonds, only the virtual orbital exhibits electron density at the N atoms. The virtual orbitals in the A\textsubscript{u} state include the 1a\textsubscript{u} and 2a\textsubscript{u} orbitals, which contribute to the signal at small q around the N atoms and at large q around the C atoms. The virtual orbitals in the B\textsubscript{3u} state include the 2b\textsubscript{3u} and 3b\textsubscript{3u} orbitals, resulting in signal at large q around the C-C bonds and N atoms.    

%%% CONCLUSIONS
In summary, we implemented a method combining X-ray scattering theory and trajectory surface hopping approach to resolve structural changes associated with electronic transitions for polyatomic molecules. The method is demonstrated for the simulation of the averaged molecular form factor difference using pyrazine as a prototypical molecule. The calculated signals directly reflect the wavepacket motion in electronic excited states as well as structural changes during electronic transition for pyrazine molecule. Our method can be combined with the machine learning technique developed in Ref.~\cite{SebastianJPCL2024} to simulate time-resolved XRD of larger and more complex molecular systems. In addition, similar methods can be also used for the simulation of ultrafast electron diffraction experiments.

\section{Supporting Information}
Surface hopping protocol, computational details, additional signals for the averaged molecular form factor difference, and excited states and molecular orbitals in pyrazine.

\begin{acknowledgement}
S.V.P., J.Z. and L.P.C. acknowledge support from the starting grant of research center of new materials computing of Zhejiang Lab (No. 3700-32601). 
H.-G.D. acknowledges support from the National Natural Science Foundation of China (No.~12274247).
M.F.G. acknowledges support from the National Natural Science Foundation of China (No.~22373028). 

\end{acknowledgement}
\section{Data availability statement}
The data is available from the authors upon reasonable request.
\section{Code availability statement}

The code used to generate the spectra is available under \url{https://github.com/psebastianzjl/density_form_factor}.
\section{Competing interests}
The authors declare that they have no competing interests.
\bibliography{main}

\providecommand{\latin}[1]{#1}
\makeatletter
\providecommand{\doi}
  {\begingroup\let\do\@makeother\dospecials
  \catcode`\{=1 \catcode`\}=2 \doi@aux}
\providecommand{\doi@aux}[1]{\endgroup\texttt{#1}}
\makeatother
\providecommand*\mcitethebibliography{\thebibliography}
\csname @ifundefined\endcsname{endmcitethebibliography}  {\let\endmcitethebibliography\endthebibliography}{}
\begin{mcitethebibliography}{44}
\providecommand*\natexlab[1]{#1}
\providecommand*\mciteSetBstSublistMode[1]{}
\providecommand*\mciteSetBstMaxWidthForm[2]{}
\providecommand*\mciteBstWouldAddEndPuncttrue
  {\def\EndOfBibitem{\unskip.}}
\providecommand*\mciteBstWouldAddEndPunctfalse
  {\let\EndOfBibitem\relax}
\providecommand*\mciteSetBstMidEndSepPunct[3]{}
\providecommand*\mciteSetBstSublistLabelBeginEnd[3]{}
\providecommand*\EndOfBibitem{}
\mciteSetBstSublistMode{f}
\mciteSetBstMaxWidthForm{subitem}{(\alph{mcitesubitemcount})}
\mciteSetBstSublistLabelBeginEnd
  {\mcitemaxwidthsubitemform\space}
  {\relax}
  {\relax}

\bibitem[Kowalewski \latin{et~al.}(2017)Kowalewski, Fingerhut, Dorfman, Bennett, and Mukamel]{Mukamel2017Review}
Kowalewski,~M.; Fingerhut,~B.~P.; Dorfman,~K.~E.; Bennett,~K.; Mukamel,~S. Simulating Coherent Multidimensional Spectroscopy of Nonadiabatic Molecular Processes: From the Infrared to the X-ray Regime. \emph{Chem. Rev.} \textbf{2017}, \emph{117}, 12165--12226\relax
\mciteBstWouldAddEndPuncttrue
\mciteSetBstMidEndSepPunct{\mcitedefaultmidpunct}
{\mcitedefaultendpunct}{\mcitedefaultseppunct}\relax
\EndOfBibitem
\bibitem[Ischenko \latin{et~al.}(2017)Ischenko, Weber, and Miller]{UED1}
Ischenko,~A.~A.; Weber,~P.~M.; Miller,~R. J.~D. Capturing Chemistry in Action with Electrons: Realization of Atomically Resolved Reaction Dynamics. \emph{Chem. Rev.} \textbf{2017}, \emph{117}, 11066--11124\relax
\mciteBstWouldAddEndPuncttrue
\mciteSetBstMidEndSepPunct{\mcitedefaultmidpunct}
{\mcitedefaultendpunct}{\mcitedefaultseppunct}\relax
\EndOfBibitem
\bibitem[Young \latin{et~al.}(2018)Young, Ueda, Gühr, Bucksbaum, Simon, Mukamel, Rohringer, Prince, Masciovecchio, Meyer, Rudenko, Rolles, Bostedt, Fuchs, Reis, Santra, Kapteyn, Murnane, Ibrahim, Légaré, Vrakking, Isinger, Kroon, Gisselbrecht, L’Huillier, Wörner, and Leone]{Young_2018}
Young,~L. \latin{et~al.}  Roadmap of ultrafast x-ray atomic and molecular physics. \emph{J. Phys. B: At. Mol. Opt. Phys.} \textbf{2018}, \emph{51}, 032003\relax
\mciteBstWouldAddEndPuncttrue
\mciteSetBstMidEndSepPunct{\mcitedefaultmidpunct}
{\mcitedefaultendpunct}{\mcitedefaultseppunct}\relax
\EndOfBibitem
\bibitem[Wang \latin{et~al.}(2024)Wang, Yun, and Yang]{mol_movies_Wang2024}
Wang,~Q.; Yun,~L.; Yang,~J. Ultrafast Molecular Movies: Probing Chemical Dynamics with Femtosecond Electron and X-Ray Diffraction. \emph{CCS Chem.} \textbf{2024}, \emph{6}, 1092--1109\relax
\mciteBstWouldAddEndPuncttrue
\mciteSetBstMidEndSepPunct{\mcitedefaultmidpunct}
{\mcitedefaultendpunct}{\mcitedefaultseppunct}\relax
\EndOfBibitem
\bibitem[Neville \latin{et~al.}(2016)Neville, Averbukh, Patchkovskii, Ruberti, Yun, Chergui, Stolow, and Schuurman]{Xray1}
Neville,~S.~P.; Averbukh,~V.; Patchkovskii,~S.; Ruberti,~M.; Yun,~R.; Chergui,~M.; Stolow,~A.; Schuurman,~M.~S. Beyond structure: ultrafast X-ray absorption spectroscopy as a probe of non-adiabatic wavepacket dynamics. \emph{Faraday Discuss.} \textbf{2016}, \emph{194}, 117--145\relax
\mciteBstWouldAddEndPuncttrue
\mciteSetBstMidEndSepPunct{\mcitedefaultmidpunct}
{\mcitedefaultendpunct}{\mcitedefaultseppunct}\relax
\EndOfBibitem
\bibitem[K\"upper \latin{et~al.}(2014)K\"upper, Stern, Holmegaard, Filsinger, Rouz\'ee, Rudenko, Johnsson, Martin, Adolph, Aquila, Bajt, Barty, Bostedt, Bozek, Caleman, Coffee, Coppola, Delmas, Epp, Erk, Foucar, Gorkhover, Gumprecht, Hartmann, Hartmann, Hauser, Holl, H\"omke, Kimmel, Krasniqi, K\"uhnel, Maurer, Messerschmidt, Moshammer, Reich, Rudek, Santra, Schlichting, Schmidt, Schorb, Schulz, Soltau, Spence, Starodub, Str\"uder, Th\o{}gersen, Vrakking, Weidenspointner, White, Wunderer, Meijer, Ullrich, Stapelfeldt, Rolles, and Chapman]{Xray_isolated_kuepper}
K\"upper,~J. \latin{et~al.}  X-Ray Diffraction from Isolated and Strongly Aligned Gas-Phase Molecules with a Free-Electron Laser. \emph{Phys. Rev. Lett.} \textbf{2014}, \emph{112}, 083002\relax
\mciteBstWouldAddEndPuncttrue
\mciteSetBstMidEndSepPunct{\mcitedefaultmidpunct}
{\mcitedefaultendpunct}{\mcitedefaultseppunct}\relax
\EndOfBibitem
\bibitem[Minitti \latin{et~al.}(2015)Minitti, Budarz, Kirrander, Robinson, Ratner, Lane, Zhu, Glownia, Kozina, Lemke, Sikorski, Feng, Nelson, Saita, Stankus, Northey, Hastings, and Weber]{Imaging_minitti}
Minitti,~M.~P. \latin{et~al.}  Imaging Molecular Motion: Femtosecond X-Ray Scattering of an Electrocyclic Chemical Reaction. \emph{Phys. Rev. Lett.} \textbf{2015}, \emph{114}, 255501\relax
\mciteBstWouldAddEndPuncttrue
\mciteSetBstMidEndSepPunct{\mcitedefaultmidpunct}
{\mcitedefaultendpunct}{\mcitedefaultseppunct}\relax
\EndOfBibitem
\bibitem[Stankus \latin{et~al.}(2019)Stankus, Yong, Zotev, Ruddock, Bellshaw, Lane, Liang, Boutet, Carbajo, Robinson, \latin{et~al.} others]{stankus2019ultrafast}
Stankus,~B.; Yong,~H.; Zotev,~N.; Ruddock,~J.~M.; Bellshaw,~D.; Lane,~T.~J.; Liang,~M.; Boutet,~S.; Carbajo,~S.; Robinson,~J.~S.; others Ultrafast X-ray scattering reveals vibrational coherence following Rydberg excitation. \emph{Nat. Chem.} \textbf{2019}, \emph{11}, 716--721\relax
\mciteBstWouldAddEndPuncttrue
\mciteSetBstMidEndSepPunct{\mcitedefaultmidpunct}
{\mcitedefaultendpunct}{\mcitedefaultseppunct}\relax
\EndOfBibitem
\bibitem[Yong \latin{et~al.}(2021)Yong, Xu, Ruddock, Stankus, Carrascosa, Zotev, Bellshaw, Du, Goff, Chang, Boutet, Carbajo, Koglin, Liang, Robinson, Kirrander, Minitti, and Weber]{ultrafast_xray_yong2021}
Yong,~H. \latin{et~al.}  Ultrafast X-ray scattering offers a structural view of excited-state charge transfer. \emph{Proc. Nat. Acad. Sci. U.S.A} \textbf{2021}, \emph{118}, e2021714118\relax
\mciteBstWouldAddEndPuncttrue
\mciteSetBstMidEndSepPunct{\mcitedefaultmidpunct}
{\mcitedefaultendpunct}{\mcitedefaultseppunct}\relax
\EndOfBibitem
\bibitem[Als-Nielsen and McMorrow(2011)Als-Nielsen, and McMorrow]{als2011elements}
Als-Nielsen,~J.; McMorrow,~D. \emph{Elements of modern X-ray physics}; John Wiley \& Sons, 2011\relax
\mciteBstWouldAddEndPuncttrue
\mciteSetBstMidEndSepPunct{\mcitedefaultmidpunct}
{\mcitedefaultendpunct}{\mcitedefaultseppunct}\relax
\EndOfBibitem
\bibitem[Prince(2004)]{prince2004international}
Prince,~E. \emph{International Tables for Crystallography, Volume C: Mathematical, physical and chemical tables}; Springer Science \& Business Media, 2004\relax
\mciteBstWouldAddEndPuncttrue
\mciteSetBstMidEndSepPunct{\mcitedefaultmidpunct}
{\mcitedefaultendpunct}{\mcitedefaultseppunct}\relax
\EndOfBibitem
\bibitem[Stewart \latin{et~al.}(1975)Stewart, Bentley, and Goodman]{generalized_xrd_factors}
Stewart,~R.~F.; Bentley,~J.; Goodman,~B. {Generalized x‐ray scattering factors in diatomic molecules}. \emph{J. Chem. Phys.} \textbf{1975}, \emph{63}, 3786--3793\relax
\mciteBstWouldAddEndPuncttrue
\mciteSetBstMidEndSepPunct{\mcitedefaultmidpunct}
{\mcitedefaultendpunct}{\mcitedefaultseppunct}\relax
\EndOfBibitem
\bibitem[Bentley and Stewart(1975)Bentley, and Stewart]{stewart_diatomic_1975}
Bentley,~J.; Stewart,~R.~F. {Diatomic generalized x‐ray scattering factors: Results from Hartree–Fock electron density functions}. \emph{J. Chem. Phys.} \textbf{1975}, \emph{63}, 3794--3803\relax
\mciteBstWouldAddEndPuncttrue
\mciteSetBstMidEndSepPunct{\mcitedefaultmidpunct}
{\mcitedefaultendpunct}{\mcitedefaultseppunct}\relax
\EndOfBibitem
\bibitem[Northey \latin{et~al.}(2014)Northey, Zotev, and Kirrander]{Kirrander2014}
Northey,~T.; Zotev,~N.; Kirrander,~A. \textit{Ab Initio} Calculation of Molecular Diffraction. \emph{J. Chem. Theo. Comput.} \textbf{2014}, \emph{10}, 4911--4920\relax
\mciteBstWouldAddEndPuncttrue
\mciteSetBstMidEndSepPunct{\mcitedefaultmidpunct}
{\mcitedefaultendpunct}{\mcitedefaultseppunct}\relax
\EndOfBibitem
\bibitem[Adam \latin{et~al.}(2016)Adam, Kenichiro, and Dmitrii~V]{Kirrander2016JCTC}
Adam,~K.; Kenichiro,~S.; Dmitrii~V,~S. Ultrafast X‑ray Scattering from Molecules. \emph{J. Chem. Theory Comput.} \textbf{2016}, \emph{12}, 957--967\relax
\mciteBstWouldAddEndPuncttrue
\mciteSetBstMidEndSepPunct{\mcitedefaultmidpunct}
{\mcitedefaultendpunct}{\mcitedefaultseppunct}\relax
\EndOfBibitem
\bibitem[Daniel and et. al.(2021)Daniel, and et. al.]{MukamelPNAS2021}
Daniel,~K.; et. al. Imaging conical intersection dynamics during azobenzene photoisomerization by ultrafast X-ray diffraction. \emph{Proc. Natl. Acad. Sci. U.S.A} \textbf{2021}, \emph{118}, e2022037118\relax
\mciteBstWouldAddEndPuncttrue
\mciteSetBstMidEndSepPunct{\mcitedefaultmidpunct}
{\mcitedefaultendpunct}{\mcitedefaultseppunct}\relax
\EndOfBibitem
\bibitem[Ye \latin{et~al.}(2019)Ye, Rouxel, Cho, and Mukamel]{mukamel_xrd2}
Ye,~L.; Rouxel,~J.~R.; Cho,~D.; Mukamel,~S. Imaging electron-density fluctuations by multidimensional X-ray photon-coincidence diffraction. \emph{Proc. Nat. Acad. Sci. U.S.A} \textbf{2019}, \emph{116}, 395--400\relax
\mciteBstWouldAddEndPuncttrue
\mciteSetBstMidEndSepPunct{\mcitedefaultmidpunct}
{\mcitedefaultendpunct}{\mcitedefaultseppunct}\relax
\EndOfBibitem
\bibitem[Woywod \latin{et~al.}(1994)Woywod, Domcke, Sobolewski, and Werner]{woywod_pyrazine}
Woywod,~C.; Domcke,~W.; Sobolewski,~A.~L.; Werner,~H.-J. Characterization of the $\mathrm{S_1}$--$\mathrm{S_2}$ Conical Intersection in Pyrazine Using \textit{ab initio} Multiconfiguration Self-Consistent-Field and Multireference Configuration-Interaction Methods. \emph{J. Chem. Phys.} \textbf{1994}, \emph{100}, 1400--1413\relax
\mciteBstWouldAddEndPuncttrue
\mciteSetBstMidEndSepPunct{\mcitedefaultmidpunct}
{\mcitedefaultendpunct}{\mcitedefaultseppunct}\relax
\EndOfBibitem
\bibitem[Sala \latin{et~al.}(2014)Sala, Lasorne, Gatti, and Gu{\'e}rin]{gatti_pyrazine}
Sala,~M.; Lasorne,~B.; Gatti,~F.; Gu{\'e}rin,~S. The Role of the Low-lying Dark n$\pi$* States in the Photophysics of Pyrazine: a Quantum Dynamics Study. \emph{Phys. Chem. Chem. Phys.} \textbf{2014}, \emph{16}, 15957--15967\relax
\mciteBstWouldAddEndPuncttrue
\mciteSetBstMidEndSepPunct{\mcitedefaultmidpunct}
{\mcitedefaultendpunct}{\mcitedefaultseppunct}\relax
\EndOfBibitem
\bibitem[Xie \latin{et~al.}(2019)Xie, Sapunar, Do{\v{s}}li{\'c}, Sala, and Domcke]{xie_pyrazine}
Xie,~W.; Sapunar,~M.; Do{\v{s}}li{\'c},~N.; Sala,~M.; Domcke,~W. Assessing the Performance of Trajectory Surface Hopping Methods: Ultrafast Internal Conversion in Pyrazine. \emph{J. Chem. Phys.} \textbf{2019}, \emph{150}, 154119\relax
\mciteBstWouldAddEndPuncttrue
\mciteSetBstMidEndSepPunct{\mcitedefaultmidpunct}
{\mcitedefaultendpunct}{\mcitedefaultseppunct}\relax
\EndOfBibitem
\bibitem[Chen \latin{et~al.}(2019)Chen, Gelin, and Domcke]{CLP}
Chen,~L.; Gelin,~M.~F.; Domcke,~W. Multimode Quantum Dynamics with Multiple Davydov D2 Trial States: Application to a 24-Dimensional Conical Intersection Model. \emph{J. Chem. Phys.} \textbf{2019}, \emph{150}, 024101\relax
\mciteBstWouldAddEndPuncttrue
\mciteSetBstMidEndSepPunct{\mcitedefaultmidpunct}
{\mcitedefaultendpunct}{\mcitedefaultseppunct}\relax
\EndOfBibitem
\bibitem[Stert \latin{et~al.}(2000)Stert, Farmanara, and Radloff]{radlof_pyr_phelsp}
Stert,~V.; Farmanara,~P.; Radloff,~W. Electron Configuration Changes in Excited Pyrazine Molecules Analyzed by Femtosecond Time-Resolved Photoelectron Spectroscopy. \emph{J. Chem. Phys.} \textbf{2000}, \emph{112}, 4460--4464\relax
\mciteBstWouldAddEndPuncttrue
\mciteSetBstMidEndSepPunct{\mcitedefaultmidpunct}
{\mcitedefaultendpunct}{\mcitedefaultseppunct}\relax
\EndOfBibitem
\bibitem[Suzuki \latin{et~al.}(2010)Suzuki, Fuji, Horio, and Suzuki]{suzuki2010time}
Suzuki,~Y.~I.; Fuji,~T.; Horio,~T.; Suzuki,~T. Time-Resolved Photoelectron Imaging of Ultrafast S\textsubscript{2}→ S\textsubscript{1} Internal Conversion through Conical Intersection in Pyrazine. \emph{J. Chem. Phys.} \textbf{2010}, \emph{132}, 174302\relax
\mciteBstWouldAddEndPuncttrue
\mciteSetBstMidEndSepPunct{\mcitedefaultmidpunct}
{\mcitedefaultendpunct}{\mcitedefaultseppunct}\relax
\EndOfBibitem
\bibitem[Horio \latin{et~al.}(2016)Horio, Spesyvtsev, Nagashima, Ingle, Suzuki, and Suzuki]{suzuki2016full}
Horio,~T.; Spesyvtsev,~R.; Nagashima,~K.; Ingle,~R.~A.; Suzuki,~Y.~I.; Suzuki,~T. Full Observation of Ultrafast Cascaded Radiationless Transitions from S\textsubscript{2} ($\pi{\pi}^{*}$) State of Pyrazine using Vacuum Ultraviolet Photoelectron Imaging. \emph{J. Chem. Phys.} \textbf{2016}, \emph{145}, 044306\relax
\mciteBstWouldAddEndPuncttrue
\mciteSetBstMidEndSepPunct{\mcitedefaultmidpunct}
{\mcitedefaultendpunct}{\mcitedefaultseppunct}\relax
\EndOfBibitem
\bibitem[Gelin \latin{et~al.}(2021)Gelin, Huang, Xie, Chen, Do{\v{s}}li{\'c}, and Domcke]{OTFDW1}
Gelin,~M.~F.; Huang,~X.; Xie,~W.; Chen,~L.; Do{\v{s}}li{\'c},~N.; Domcke,~W. Ab Initio Surface-Hopping Simulation of Femtosecond Transient-Absorption Pump–Probe Signals of Nonadiabatic Excited-State Dynamics Using the Doorway–Window Representation. \emph{J. Chem. Theory Comput.} \textbf{2021}, \emph{17}, 2394--2408\relax
\mciteBstWouldAddEndPuncttrue
\mciteSetBstMidEndSepPunct{\mcitedefaultmidpunct}
{\mcitedefaultendpunct}{\mcitedefaultseppunct}\relax
\EndOfBibitem
\bibitem[Huang \latin{et~al.}(2021)Huang, Xie, Do{\v{s}}li{\'c}, Gelin, and Domcke]{Xiang2D}
Huang,~X.; Xie,~W.; Do{\v{s}}li{\'c},~N.; Gelin,~M.~F.; Domcke,~W. Ab Initio Quasiclassical Simulation of Femtosecond Time-Resolved Two-Dimensional Electronic Spectra of Pyrazine. \emph{J. Phys. Chem. Lett.} \textbf{2021}, \emph{12}, 11736--11744\relax
\mciteBstWouldAddEndPuncttrue
\mciteSetBstMidEndSepPunct{\mcitedefaultmidpunct}
{\mcitedefaultendpunct}{\mcitedefaultseppunct}\relax
\EndOfBibitem
\bibitem[Sun \latin{et~al.}(2020)Sun, Xie, Chen, Domcke, and Gelin]{Skw20a}
Sun,~K.; Xie,~W.; Chen,~L.; Domcke,~W.; Gelin,~M.~F. Multi-Faceted Spectroscopic Mapping of Ultrafast Nonadiabatic Dynamics near Conical Intersections: A Computational Study. \emph{J. Chem. Phys.} \textbf{2020}, \emph{153}, 174111\relax
\mciteBstWouldAddEndPuncttrue
\mciteSetBstMidEndSepPunct{\mcitedefaultmidpunct}
{\mcitedefaultendpunct}{\mcitedefaultseppunct}\relax
\EndOfBibitem
\bibitem[Chen \latin{et~al.}(2021)Chen, Sun, Shalashilin, Gelin, and Zhao]{LP21}
Chen,~L.; Sun,~K.; Shalashilin,~D.~V.; Gelin,~M.~F.; Zhao,~Y. Efficient Simulation of Time- and Frequency-Resolved Four-Wave-Mixing Signals with a Multiconfigurational Ehrenfest Approach. \emph{J. Chem. Phys.} \textbf{2021}, \emph{154}, 054105\relax
\mciteBstWouldAddEndPuncttrue
\mciteSetBstMidEndSepPunct{\mcitedefaultmidpunct}
{\mcitedefaultendpunct}{\mcitedefaultseppunct}\relax
\EndOfBibitem
\bibitem[Pite{\v{s}}a \latin{et~al.}(2021)Pite{\v{s}}a, Sapunar, Ponzi, Gelin, Do{\v{s}}li{\'c}, Domcke, and Decleva]{OTFDW2}
Pite{\v{s}}a,~T.; Sapunar,~M.; Ponzi,~A.; Gelin,~M.~F.; Do{\v{s}}li{\'c},~N.; Domcke,~W.; Decleva,~P. Combined Surface-Hopping, Dyson Orbital, and B-Spline Approach for the Computation of Time-Resolved Photoelectron Spectroscopy Signals: The Internal Conversion in Pyrazine. \emph{J. Chem. Theory Comput.} \textbf{2021}, \emph{17}, 5098--5109\relax
\mciteBstWouldAddEndPuncttrue
\mciteSetBstMidEndSepPunct{\mcitedefaultmidpunct}
{\mcitedefaultendpunct}{\mcitedefaultseppunct}\relax
\EndOfBibitem
\bibitem[Kaczun \latin{et~al.}(2023)Kaczun, Dempwolff, Huang, Gelin, Domcke, and Dreuw]{OTFDW4}
Kaczun,~T.; Dempwolff,~A.~L.; Huang,~X.; Gelin,~M.~F.; Domcke,~W.; Dreuw,~A. Tuning UV Pump X-ray Probe Spectroscopy on the Nitrogen K Edge Reveals the Radiationless Relaxation of Pyrazine: Ab Initio Simulations Using the Quasiclassical Doorway--Window Approximation. \emph{J. Phys. Chem. Lett.} \textbf{2023}, \emph{14}, 5648--5656\relax
\mciteBstWouldAddEndPuncttrue
\mciteSetBstMidEndSepPunct{\mcitedefaultmidpunct}
{\mcitedefaultendpunct}{\mcitedefaultseppunct}\relax
\EndOfBibitem
\bibitem[Freibert \latin{et~al.}(2022)Freibert, Mendive-Tapia, Huse, and Vendrell]{Freibert_2021}
Freibert,~A.; Mendive-Tapia,~D.; Huse,~N.; Vendrell,~O. Femtosecond x-ray absorption spectroscopy of pyrazine at the nitrogen K-edge: on the validity of the Lorentzian limit. \emph{J. Phys. B: At. Mol. Opt. Phys.} \textbf{2022}, \emph{54}, 244003\relax
\mciteBstWouldAddEndPuncttrue
\mciteSetBstMidEndSepPunct{\mcitedefaultmidpunct}
{\mcitedefaultendpunct}{\mcitedefaultseppunct}\relax
\EndOfBibitem
\bibitem[Freibert \latin{et~al.}(2024)Freibert, Mendive-Tapia, Huse, and Vendrell]{freibert_2024}
Freibert,~A.; Mendive-Tapia,~D.; Huse,~N.; Vendrell,~O. Time-Dependent Resonant Inelastic X-ray Scattering of Pyrazine at the Nitrogen K-Edge: A Quantum Dynamics Approach. \emph{J. Chem. Theory Comput.} \textbf{2024}, \emph{20}, 2167--2180\relax
\mciteBstWouldAddEndPuncttrue
\mciteSetBstMidEndSepPunct{\mcitedefaultmidpunct}
{\mcitedefaultendpunct}{\mcitedefaultseppunct}\relax
\EndOfBibitem
\bibitem[James(1982)]{james_optical_principles1982}
James,~R.~W. \emph{The optical principles of the diffraction of X-rays}; Ox Bow Press, 1982\relax
\mciteBstWouldAddEndPuncttrue
\mciteSetBstMidEndSepPunct{\mcitedefaultmidpunct}
{\mcitedefaultendpunct}{\mcitedefaultseppunct}\relax
\EndOfBibitem
\bibitem[Schülke(2007)]{schuelke_electron_dyn2007}
Schülke,~W. \emph{{Electron Dynamics by Inelastic X-ray Scattering}}; Oxford University Press, 2007\relax
\mciteBstWouldAddEndPuncttrue
\mciteSetBstMidEndSepPunct{\mcitedefaultmidpunct}
{\mcitedefaultendpunct}{\mcitedefaultseppunct}\relax
\EndOfBibitem
\bibitem[Crespo-Otero and Barbatti(2018)Crespo-Otero, and Barbatti]{OTF2}
Crespo-Otero,~R.; Barbatti,~M. Recent Advances and Perspectives on Nonadiabatic Mixed Quantum–Classical Dynamics. \emph{Chem. Rev.} \textbf{2018}, \emph{118}, 7026--7068\relax
\mciteBstWouldAddEndPuncttrue
\mciteSetBstMidEndSepPunct{\mcitedefaultmidpunct}
{\mcitedefaultendpunct}{\mcitedefaultseppunct}\relax
\EndOfBibitem
\bibitem[Nelson \latin{et~al.}(2020)Nelson, White, Bjorgaard, Sifain, Zhang, Nebgen, Fernandez-Alberti, Mozyrsky, Roitberg, and Tretiak]{OTF3}
Nelson,~T.~R.; White,~A.~J.; Bjorgaard,~J.~A.; Sifain,~A.~E.; Zhang,~Y.; Nebgen,~B.; Fernandez-Alberti,~S.; Mozyrsky,~D.; Roitberg,~A.~E.; Tretiak,~S. Non-adiabatic Excited-State Molecular Dynamics: Theory and Applications for Modeling Photophysics in Extended Molecular Materials. \emph{Chem. Rev.} \textbf{2020}, \emph{120}, 2215--2287\relax
\mciteBstWouldAddEndPuncttrue
\mciteSetBstMidEndSepPunct{\mcitedefaultmidpunct}
{\mcitedefaultendpunct}{\mcitedefaultseppunct}\relax
\EndOfBibitem
\bibitem[Curchod and Mart{\'\i}nez(2018)Curchod, and Mart{\'\i}nez]{OTF1}
Curchod,~B. F.~E.; Mart{\'\i}nez,~T.~J. Ab Initio Nonadiabatic Quantum Molecular Dynamics. \emph{Chem. Rev.} \textbf{2018}, \emph{118}, 3305--3336\relax
\mciteBstWouldAddEndPuncttrue
\mciteSetBstMidEndSepPunct{\mcitedefaultmidpunct}
{\mcitedefaultendpunct}{\mcitedefaultseppunct}\relax
\EndOfBibitem
\bibitem[Mai and Gonz{\'a}lez(2020)Mai, and Gonz{\'a}lez]{mai2020molecular}
Mai,~S.; Gonz{\'a}lez,~L. Molecular Photochemistry: Recent Developments in Theory. \emph{Angew. Chem. Int. Ed.} \textbf{2020}, \emph{59}, 16832--16846\relax
\mciteBstWouldAddEndPuncttrue
\mciteSetBstMidEndSepPunct{\mcitedefaultmidpunct}
{\mcitedefaultendpunct}{\mcitedefaultseppunct}\relax
\EndOfBibitem
\bibitem[Belyaev \latin{et~al.}(2014)Belyaev, Lasser, and Trigila]{LZM1}
Belyaev,~A.~K.; Lasser,~C.; Trigila,~G. Landau--Zener Type Surface Hopping Algorithms. \emph{J. Chem. Phys.} \textbf{2014}, \emph{140}, 224108\relax
\mciteBstWouldAddEndPuncttrue
\mciteSetBstMidEndSepPunct{\mcitedefaultmidpunct}
{\mcitedefaultendpunct}{\mcitedefaultseppunct}\relax
\EndOfBibitem
\bibitem[Belyaev \latin{et~al.}(2015)Belyaev, Domcke, Lasser, and Trigila]{LZM2}
Belyaev,~A.~K.; Domcke,~W.; Lasser,~C.; Trigila,~G. Nonadiabatic Nuclear Dynamics of the Ammonia Cation Studied by Surface Hopping Classical Trajectory Calculations. \emph{J. Chem. Phys.} \textbf{2015}, \emph{142}, 104307\relax
\mciteBstWouldAddEndPuncttrue
\mciteSetBstMidEndSepPunct{\mcitedefaultmidpunct}
{\mcitedefaultendpunct}{\mcitedefaultseppunct}\relax
\EndOfBibitem
\bibitem[Suchan \latin{et~al.}(2020)Suchan, Jano{\v{s}}, and Slaví{\v{c}}ek]{pragmatic_lz}
Suchan,~J.; Jano{\v{s}},~J.; Slaví{\v{c}}ek,~P. Pragmatic Approach to Photodynamics: Mixed Landau–Zener Surface Hopping with Intersystem Crossing. \emph{J. Chem. Theory Comput.} \textbf{2020}, \emph{16}, 5809--5820\relax
\mciteBstWouldAddEndPuncttrue
\mciteSetBstMidEndSepPunct{\mcitedefaultmidpunct}
{\mcitedefaultendpunct}{\mcitedefaultseppunct}\relax
\EndOfBibitem
\bibitem[Zhan \latin{et~al.}(2023)Zhan, Gelin, Huang, and Sun]{kewei_beating_maps_2023}
Zhan,~S.; Gelin,~M.~F.; Huang,~X.; Sun,~K. {Ab initio simulation of peak evolutions and beating maps for electronic two-dimensional signals of a polyatomic chromophore}. \emph{J. Chem. Phys.} \textbf{2023}, \emph{158}, 194106\relax
\mciteBstWouldAddEndPuncttrue
\mciteSetBstMidEndSepPunct{\mcitedefaultmidpunct}
{\mcitedefaultendpunct}{\mcitedefaultseppunct}\relax
\EndOfBibitem
\bibitem[Pios \latin{et~al.}(2024)Pios, Gelin, Ullah, Dral, and Chen.]{SebastianJPCL2024}
Pios,~S.~V.; Gelin,~M.~F.; Ullah,~A.; Dral,~P.~O.; Chen.,~L. Artificial-Intelligence-Enhanced On-the-Fly Simulation of Nonlinear Time-Resolved Spectra. \emph{J. Phys. Chem. Lett.} \textbf{2024}, \emph{15}, 2325--2331\relax
\mciteBstWouldAddEndPuncttrue
\mciteSetBstMidEndSepPunct{\mcitedefaultmidpunct}
{\mcitedefaultendpunct}{\mcitedefaultseppunct}\relax
\EndOfBibitem
\end{mcitethebibliography}
\end{document}